\documentclass[aps,prl,superscriptaddress,twocolumn,showpacs,preprintnumbers,amsmath,amssymb]{revtex4-1}
\usepackage{graphicx}
\usepackage{dcolumn}
\usepackage{bm}
\usepackage{upgreek}

\begin{document}

\title{The anti-ordinary Hall effect in NiPt thin films}
\author{Taras Golod}
\affiliation{Department of Physics, Stockholm University, AlbaNova University Center, SE-10691 Stockholm, Sweden}
\author{Andreas Rydh}
\affiliation{Department of Physics, Stockholm University, AlbaNova University Center, SE-10691 Stockholm, Sweden}

\author{Peter Svedlindh}
\affiliation{Department of Engineering Sciences, Uppsala University, Box 534, SE-75121 Uppsala, Sweden}

\author{Vladimir M. Krasnov}
\email{Vladimir.Krasnov@fysik.su.se}
\affiliation{Department of Physics, Stockholm University, AlbaNova University Center, SE-10691 Stockholm, Sweden}

\date{\today }

\begin{abstract}
We study the anomalous Hall effect in binary alloys between the
group-10 elements Ni and Pt. It is observed that the ordinary Hall
effect is negative (electron-like) at any composition of the
alloy. The extraordinary Hall effect is also negative except in
the vicinity of the ferromagnetic quantum critical point. Close to
the critical point the sign of the extraordinary Hall effect can
be changed to positive (hole-like) by tuning either the
temperature or the composition of the alloy. We attribute such an
``anti-ordinary" Hall effect with opposite signs of the ordinary
and the extraordinary contributions to a Berry phase singularity,
moving away from the Fermi energy with increasing the
ferromagnetic exchange energy.
\end{abstract}




\maketitle

Hall effect in ferromagnetic metals comprises the ordinary and the
extraordinary contributions, proportional to the applied magnetic
field $H$ and the magnetization $M$, respectively \cite{Hurd}:
\begin{equation}\label{Rxy_AHE}
\rho_{xy}=R_0 H + R_1 M.
\end{equation}
Here $R_0$ and $R_1$ are the ordinary and the extraordinary Hall
coefficients.  The ordinary Hall effect (OHE) is caused by the
action of the spin-independent Lorentz force on itinerant charge
carriers. The sign of the OHE reflects the topology of the Fermi
surface \cite{ARPES}, it is negative for electron-like and
positive for hole-like charge carriers. The extraordinary Hall
effect (EHE) arises from various spin-orbit interactions
\cite{Hurd}. However, the mechanisms of the extraordinary Hall
effect (EHE) are still under debate
\cite{Nagaosa,OrbitHall,Souza2007,Ebert2011,Mokrousov2011}, making
the EHE possibly the oldest unsolved problem in solid state
physics. Identification of the EHE mechanisms is obscured by the
coexistence of intrinsic (electron-structure related)
\cite{Karplus1954} and extrinsic (impurity and scattering related)
\cite{Skew1955,Berger1970,Kovalev,MokrousovPRL2011} contributions
that are difficult to disentangle.

NiPt alloys are ideal for analysis of the EHE: (i) Ni and Pt have
a perfect chemical matching. They belong to the same group-10 in
the periodic table, have similar electronic
[$(n-1)$d$^9$$n$s$^1$], and crystal (fcc) structures and form
solid solutions at any concentrations (unlike many binary alloys
that are prone to phase segregation \cite{Hurd,Abrikosov}). (ii)
Pt is characterized by large spin-orbit coupling, needed for the
EHE. (iii) Ni has a sharp singularity in the minority d-band near
the Fermi surface \cite{Kotliar2001}. Both latter factors amplify
the intrinsic EHE in NiPt alloys \cite{Souza2007,Mokrousov2011}.
(iv) A binary alloy between a magnetic (Ni) and a normal (Pt)
metal is the simplest system in which spin-splitting of the
electronic system and the ferromagnetic Curie temperature $T_C$
can be continuously tuned by changing the alloy composition. For
Ni$_x$Pt$_{1-x}$ alloys ferromagnetism appears at the critical Ni
concentration $x_c\simeq 0.4$ \cite{Golod2011}. The fact that
$T_C=0$ at $x=x_c$ indicates the occurrence of a ferromagnetic
quantum phase transition (QPT), driven by quantum rather than
thermal fluctuations \cite{QPT_Rev2007}.

Here we study Hall effect in NiPt thin films. It is observed that
the OHE remains negative (electron-like) at any composition of the
alloy. The EHE is also negative except in the vicinity of the
ferromagnetic quantum critical point. Close to the critical point
the sign of the extraordinary Hall effect can be changed to
positive (hole-like) by tuning either the temperature or the
composition of the alloy. We attribute such an ``anti-ordinary"
Hall effect with opposite signs of the OHE and the EHE to a Berry
phase singularity, shifting away from the Fermi energy with
increasing ferromagnetic exchange energy. The spectroscopic
information contained in the observed anti-ordinary Hall effect
indicates its intrinsic (electron-structure related) nature.


\begin{figure*}[t]
    \centering
    \includegraphics[width=\textwidth]{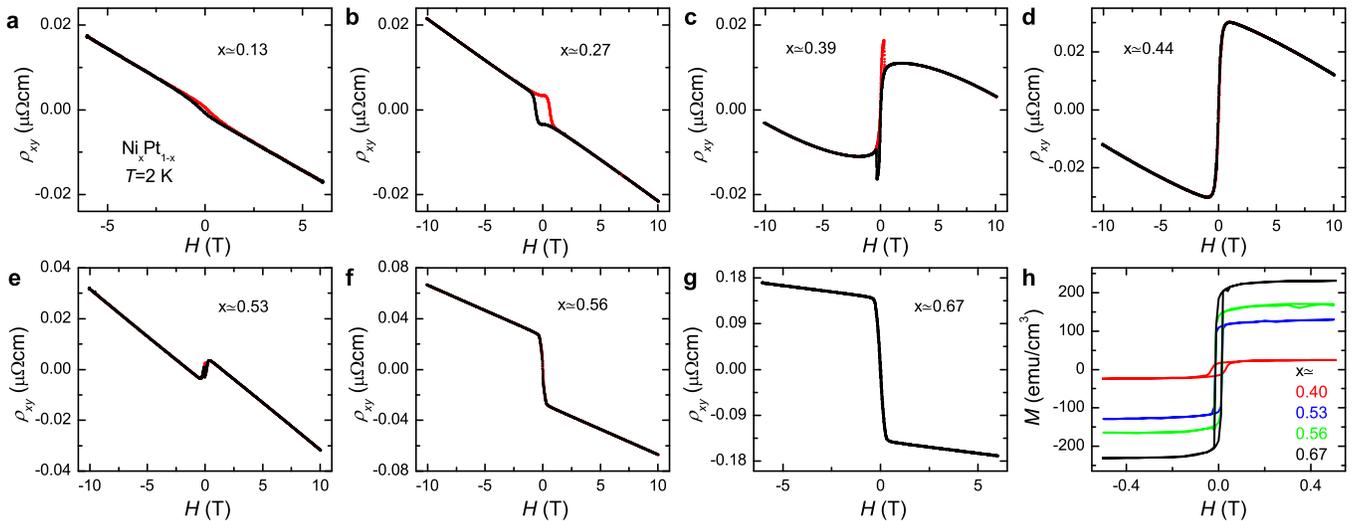}
    \caption{(Color online). 
    \textbf{a}-\textbf{g}, Hall resistivity of Ni$_x$Pt$_{1-x}$ thin films with different Ni
    concentrations at $T$ = 2 K for increasing (black) and decreasing (red) fields perpendicular
    to films. The linear-in-field OHE is clearly seen at large fields. The OHE is electron-like
    irrespective of concentration. The step-like change of $\rho_{xy}$ at low fields is due to
    the EHE. With increasing Ni concentration the EHE changes sign and become hole-like close to
    the critical ferromagnetic concentration $x_c\simeq 0.4$, indicating the occurrence of the anti-ordinary Hall effect. \textbf{h}, Magnetization vs. in-plane field $H$ for films with different Ni concentrations measured at $T$ = 2 K ($x\simeq 0.40$, 0.53 and 0.56) and 10 K ($x\simeq 0.67$).}
    \label{fig:fig1}
\end{figure*}


NiPt thin films with thicknesses $\sim 35-45$ nm were made by
magnetron co-sputtering on oxidized Si wafers at room temperature.
The compositions of the films were determined by energy-dispersive
X-ray spectroscopy \cite{Golod2011}. The films were patterned into
Hall bridges using photolithography and ion milling. Resistive
measurements were made in a $^4$He-gas flow cryostat with a
superconducting solenoid. All Hall measurements were made in field
perpendicular to the film. Magnetization measurements were made
using a commercial SQUID magnetometer (Quantum Design MPMS 5XL).
To avoid complications with the demagnetization factor, the
magnetization was measured with the field parallel to the films.
More details on sample fabrication, characterization and the
experimental setup can be found in Ref. \cite{Golod2011}.

Figure~1 a-g shows Hall resistivities $\rho_{xy} (H)$ of NiPt thin
films with different concentrations at $T=2$ K. The
linear-in-field OHE contribution is clearly seen at large fields.
It changes only modestly with concentration. Note that for all
alloys $R_0=d\rho_{xy}/dH$ is negative, which is also the case for
pure Ni and Pt \cite{Hurd}. The step-like change of $\rho_{xy}$ at
low fields is due to the EHE and reflects the step-like change of
the magnetization $M(H)$, shown in Fig.~1h. It is seen that at low
Ni concentrations, Fig.~1a,b, the EHE has the same sign as the OHE
(at low $T$ such films are in a frozen cluster-glass state with a
perpendicular magnetic anisotropy, leading to a relatively large
hysteresis in $\rho_{xy} (H)$ \cite{Golod2011}). Upon approaching
the critical ferromagnetic concentration $x_c \simeq 0.4$,
Fig.~1c,d, the EHE changes sign and becomes positive, indicating
the occurrence of the anti-ordinary Hall effect
\cite{Mokrousov2011}. With further increase of Ni concentration
the positive EHE decreases, almost quenches at $x\simeq 0.53$ and
then again changes sign and becomes negative in Ni-rich alloys,
Fig.~1f,g.

Figure~2 shows the $T$-variation of the $\rho_{xy} (H)$ curves for
different films. In films with low Ni concentration, Fig.~2a, both
the OHE and the EHE are negative at all temperatures. Close to the
critical Ni concentration $x_c\simeq 0.4$ the EHE becomes positive
at low $T = 2$ K, as seen from Fig.~2b, indicating the appearance
of the anti-ordinary Hall effect.
At a higher Ni concentration $x\simeq 0.44$, Fig.~2c, the EHE
has a large positive value at low $T=2$ K. With increasing $T$ the
positive EHE decreases, acquiring a small negative value before
vanishing at high $T$. At $x\simeq 0.53$, Fig.~2d, the EHE at
low $T=2$ K is almost quenched, but the large positive EHE is
restored at elevated $T=100$ K. At higher Ni concentration
$x\simeq 0.56$, Fig.~2e, the EHE at low $T$ becomes negative,
but the magnitude of the negative EHE decreases with increasing temperature and a small positive EHE reemerges at $T=150$ K. Finally at
large Ni concentration $x\simeq 0.67$, Fig.~2f, the anti-ordinary Hall
effect disappears. Here the EHE is again negative at all $T$ and
vanishes upon approaching $T_C$.

Figure~3 summarizes composition dependencies of NiPt film
characteristics. Figure~3a shows the residual longitudinal
resistivity $\rho_{xx0}$ at $T=2$ K and $H=0$. It follows a
standard parabolic behavior for binary alloys (scattering of the
data is caused by a varying surface contribution due to different
film thicknesses). Figure~3b shows the OHE coefficient at $T=2$ K,
obtained from the large-field slope of the $\rho_{xy}(H)$ curves
shown in Fig.~1: $R_0 = d\rho_{xy}/dH_{[H\sim 10\,\mathrm{T}]}$.
It is seen that the OHE remains negative at all concentrations.
Figure~3c displays the saturated value of the EHE resistivity at
$T=2$ K: $\rho_{\mathrm{EHE}} = \rho_{xy} - (d\rho_{xy}/dH_{[H\sim
10\,\mathrm{T}]})H$. From Fig.~3c it is clear that the onset of
the anti-ordinary Hall effect, indicated by the arrow, appears in
the vicinity of the critical Ni concentration $x_c \simeq 0.4$.

\begin{figure*}[t]
    \centering
    \includegraphics[width=0.9\textwidth]{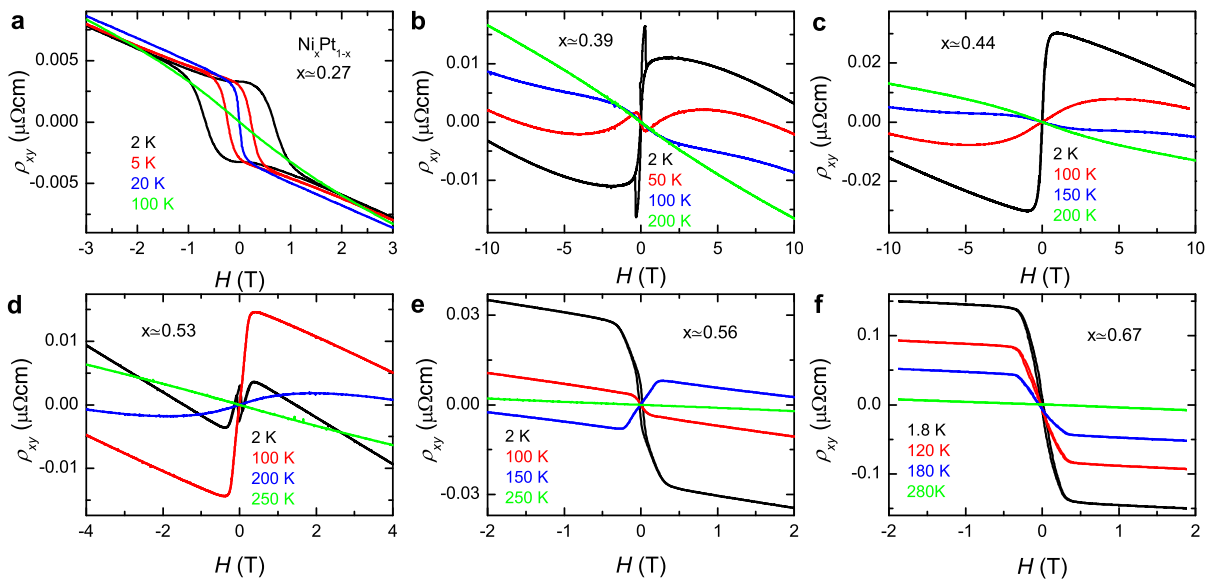}
    \caption{(Color online). 
    \textbf{a}-\textbf{f}, $T$-evolution of $\rho_{xy}$ vs. $H$ curves for films with Ni concentrations $x\simeq 0.27$ (\textbf{a}), 0.39 (\textbf{b}), 0.44 (\textbf{c}), 0.53 (\textbf{d}), 0.56 (\textbf{e}) and 0.67 (\textbf{f}). For films with low (\textbf{a}) and high (\textbf{f}) Ni concentrations both the OHE and the EHE are electron-like at all temperatures. However, in films with near critical concentration $x_c\simeq 0.4$ the EHE changes sign to become hole-like at a certain temperature (\textbf{b}-\textbf{e}), while the OHE always remains electron-like.}
    \label{fig:fig2}
\end{figure*}

In Fig.~3d,e, the temperature dependence of the
magnetization at $H=0.05$ T and the EHE resistivity at $|H|=0.16$
T are displayed. The Curie temperature for the films with
super-critical Ni concentration is well defined from the onset of
magnetization. Figure~3e demonstrates that the EHE also changes
sign as a function of temperature at concentrations close to the
critical, as seen from Fig.~2b-e.

To understand the origin of the observed sign-reversal of the EHE,
we first recollect factors that determine the sign of the Hall
effect. The sign of the OHE is determined by the sign on the
effective mass of charge carriers, reflecting the curvature of the
Fermi surface \cite{Hurd,ARPES}. From Fig. 3b it is seen that the
OHE remains electron-like at all concentrations,
indicating that there is no change of the topology for the major
part of the Fermi surface. Therefore, the discussed sign change of
the EHE in NiPt is qualitatively different from that in binary
alloys between metals with hole-like (e.g. Fe or Co) and
electron-like (e.g. Cu, Pd, Pt, Au) carriers, in which the sign
change of both OHE and EHE as a function of composition can be
associated with the change of carrier type
\cite{Hurd,Souza2007,Mokrousov2011}. In our case we are mixing two
electron-like metals from group-10 in the periodic table, with
similar crystal and band structures and for which the majority of
mobile (s-type) charge carriers remain always electron-like. It is
only the EHE that changes sign close to the ferromagnetic quantum
critical point $x=x_c$, $T_C=0$.

The sign of the extrinsic skew scattering EHE mechanism
\cite{Skew1955} may depend on the sign of the interaction with the
scatterer (attractive or repulsive).
However, the skew scattering can not explain the observed multiple
sign reversals of the EHE in our NiPt films as a function of
concentration and temperature. For a given set of elements (Ni and
Pt) the sign of the interaction should not depend on
concentrations or $T$. Furthermore, the skew mechanism is
significant only in clean materials with low resistivity
\cite{Nagaosa}. Our diluted (both Ni and Pt rich) films, with the
lowest resistivities, Fig.~3a, have the same electronic sign of
both the OHE and the EHE, Fig.~1a,b,g, i.e., the skew scattering,
if present, leads to a negative EHE. Only films with near critical
concentrations, that are characterized by fairly large
resistivities $\rho_{xx}\sim 30-40 ~\upmu \Omega \mathrm{cm}$, see
Fig.~3a, exhibit the anti-ordinary Hall effect. At those
resistivities the intrinsic mechanism is usually dominant
\cite{Nagaosa}. Furthermore, for films with concentrations
slightly larger than $x_c$ the EHE again changes sign to become
anti-ordinary with increasing temperature, Fig.~3e, i.e., with
increasing scattering rate, clearly indicating that the positive
contribution to the EHE is not due to the skew scattering.

Discrimination between the intrinsic \cite{Karplus1954} and the
extrinsic side jump \cite{Berger1970} EHE mechanisms is more
complicated. Both are scattering-independent and can be connected
to the Fermi surface topology, but in a way significantly
different from the OHE case. Namely, they are connected to the
Berry phase curvature
\cite{Nagaosa,Mokrousov2011,Souza2007,Ebert2011,Kovalev,MokrousovPRL2011},
rather than the Fermi surface curvature. As a result, the EHE for
both mechanisms arises only from singular points occupying small
parts of the Fermi surface. However, the two EHE mechanisms may
arise from different parts at the Fermi surface and may have
different signs \cite{Souza2007,MokrousovPRL2011}.
The most important difference for our case is that the amplitude
of the extrinsic side-jump EHE is only weakly dependent on
parameters of the electronic system and usually maintains the same
sign upon varying external parameters
\cite{Souza2007,MokrousovPRL2011,Ebert2011}. To the contrary, the
intrinsic EHE is very sensitive to parameters of the electronic
system and may experience significant changes both in the
amplitude and even in the sign \cite{Souza2007,SrRuO}. Therefore,
sign reversal EHE as a function of composition or temperature,
observed in a variety of materials \cite{Hurd}, is often
attributed either to a competition between extrinsic and intrinsic
contributions with different signs \cite{Gd,FeTaS,Ebert2011}, or
solely to the sign change of the intrinsic EHE
\cite{Nagaosa,SrRuO,Mokrousov2011,Tesanovic}.

\begin{figure*}[t]
    \centering
    \includegraphics[width=0.9\textwidth]{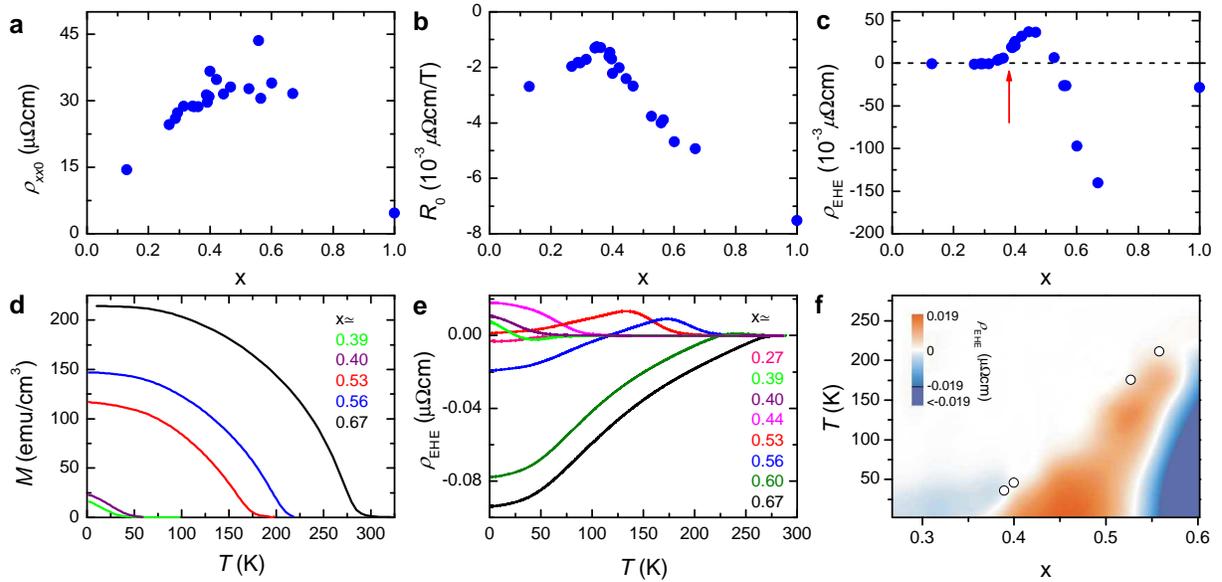}
    \caption{(Color online). 
    \textbf{a}-\textbf{c}, The residual longitudinal resistivity at $T=2$ K and $H=0$ T (\textbf{a}), the OHE coefficient $R_{0}$ at $T=2$ K (\textbf{b}) and the extraordinary Hall resistivity $\rho_{\mathrm{EHE}}$ measured at $H=10$ T and $T=2$ K (\textbf{c}). Panels \textbf{d} and \textbf{e} show temperature dependencies of magnetization at $H=0.05$ T and the extraordinary Hall resistivity at $|H|=0.16$ T for films with different compositions. \textbf{f}, The $T-x$ diagram (color plot) of the extraordinary Hall resistivity obtained from \textbf{e}. The anti-ordinary Hall effect appears in the vicinity of the quantum phase transition, $x_c\simeq 0.4$, $T_C=0$. The open circles in \textbf{f} indicate $T_C$ estimated from the onset of $M$(T) in \textbf{d}.}
    \label{fig:fig3}
\end{figure*}

As seen from Fig. 3b, the OHE for our NiPt films remains negative
at all concentrations, indicating that there is no change of the
topology for the major part of the Fermi surface. This brings us
to the conclusion that the observed multiple sign changes of the
EHE are caused by the intrinsic EHE mechanism, associated with the
Berry phase singularity, occupying a relatively small portion of
the Fermi surface. It is qualitatively similar to the explanation
of the sign change of the EHE in Sr$_{1-x}$Ca$_x$RuO$_3$ as a
function of Ca doping \cite{SrRuO}. In that case it was argued
that the sign change was simply caused by the increase of the EHE
coefficient $R_1$ with increasing magnetization. However, unlike
Sr$_{1-x}$Ca$_x$RuO$_3$, in NiPt films there is no obvious scaling
between $R_1$ and $M$ in Eq. (1). This is most clear from the
comparison of the $M(T)$, Fig.~3d, and $\rho_{\mathrm{EHE}}(T)$,
Fig.~3e, dependencies; at a similar magnetization the
near-critical films have positive EHE, while the rest of the films
have negative EHE. Clearly, there is no universal scaling
$R_1(M)$, but $R_1$ is varying with Ni concentration.

Figure~3f shows  the $T-x$ diagram of the EHE resistivity for
the NiPt films. It is clear that positive EHE emerges at the
ferromagnetic QPT, $x_c\simeq 0.4$, $T_C=0$. With increasing Ni
concentration $x>x_c$ the EHE turns negative at low $T$, but the
positive EHE is restored at an elevated $T$. Such a behavior can be
understood if the Berry phase singularity, responsible for the
sign change of the EHE, crosses the Fermi level at $x=x_c$ and
moves away from the Fermi level at $x>x_c$. Since the
singularity is lying above the chemical potential at $x>x_c$, the
anti-ordinary EHE disappears at $T\rightarrow 0$. However, it
reemerges with increasing temperature, when the singularity is
repopulated by thermally excited quasiparticles. In this case the
temperature of the maximum anti-ordinary EHE is a measure of the
energy shift of the singularity. At all concentrations the maximum
occurs slightly below $T_C$, consistent with the assumption that
the energy of the singularity is connected with the ferromagnetic
exchange energy.

To conclude, we have reported on the unusual ``anti-ordinary" Hall
effect in NiPt thin films. The phenomenon is attributed to the
Berry phase singularity in the spin-polarized d-band part of the
Fermi surface.
We are able to trace how the singularity moves away from the Fermi
level with increasing Ni concentration, in correlation with the
increase of the exchange energy. Such spectroscopic information,
contained in the anti-ordinary Hall effect clearly indicates that
it is of intrinsic, electronic structure-related, nature.

Support from the Swedish Research Council, and the SU-Core
facility in Nanotechnology is gratefully acknowledged.

\end{document}